# Using thermo-optical nonlinearity to robustly separate absorption and radiation losses in nanophotonic resonators

Mingkang Wang[1,2,†], Diego J. Perez-Morelo[1,2], Vladimir Aksyuk[1,*]

[1]Microsystems and Nanotechnology Division, National Institute of Standards and Technology, Gaithersburg, MD 20899 USA

[2]Institute for Research in Electronics and Applied Physics, University of Maryland, College Park, MD 20742, USA

Correspondence:

† Mingkang Wang, tel:+1-301-605-4531; mingkang.wang@nist.gov

*Vladimir Aksyuk, tel: +1-301-975-2867; vladimir.aksyuk@nist.gov

## Abstract

*Low-loss nanophotonic resonators have been widely used in fundamental science and applications thanks to their ability to concentrate optical energy. Key for resonator engineering, the total intrinsic loss is easily determined by spectroscopy, however, quantitatively separating absorption and radiative losses is challenging. While the concentrated heat generated by absorption within the small mode volume results in generally unwanted thermo-optical effects, they can provide a way for quantifying absorption. Here, we propose and experimentally demonstrate a technique for separating the loss mechanisms with high confidence using only linear spectroscopic measurements. We use the optically measured resonator thermal time constant to experimentally connect the easily-calculable heat capacity to the thermal impedance, needed to calculate the absorbed power from the temperature change. We report the absorption, radiation, and coupling losses for ten whispering-gallery modes of three different radial orders on a Si microdisk. Similar absorptive loss rates are found for all the modes, despite order-of-magnitude differences in the total dissipation rate due to widely differing radiation losses. Measuring radiation losses of many modes enables distinguishing the two major components of radiation loss originating from scattering and leakage. The all-optical characterization technique is applicable to any nanophotonic resonators subject to thermo-optical effects.*

## Introduction

Nanophotonic resonators can combine an ultra-high quality factor with a small mode volume. The highly concentrated energy generates heat via the absorption loss of the resonator. The resulting strong temperature variations largely distort the optical response via changing the refractive index and the size of the resonator, inducing a strong thermo-optical effect[1]. While the thermo-optical effect can be used for resonance wavelength tuning[2–4], thermal sensing[5–7], thermal locking[8,9], and thermal imaging[10–12], the heat from the light absorbed within the cavity generates undesired thermal-instability[13,14] and nonlinear dynamics[15–17], and can strongly limit sensitivity and bandwidth of nanophotonic sensors[18]. A technique for experimentally distinguishing absorption losses from radiation losses with high confidence would help guide the design and fabrication strategies for nanophotonic resonators, including lowering absorptive loss rates to reduce the thermo-optical nonlinearities, as well as quantifying and harnessing these nonlinearities.

The thermo-optical effect itself has been shown to be useful for the analysis of the loss mechanisms in nanophotonic resonators, for example, by measuring the nonlinear absorption rate[19,20], estimating cavity temperature changes based on simulation[15], and fitting the loss as a function of microdisk radius[21] or wavelength-dependent absorption coefficient[22]. However, the previously proposed methods are not generally applicable since they demand either strong material optical nonlinearities, high quality factors, or multiple devices to sweep parameter space. Finite element method (FEM) simulations unchecked by experiment may not always accurately capture thermal transport effects such as interfacial thermal impedances, radiative and gas-phase heat transport, and variations in material thermal conductivity. The previous studies so far only focused on a single optical mode or modes with the same radial number and similar quality factors.

Here, we propose a general dissipation analysis method that separates the absorption loss from other sources by calculating the amount of optical power that is being converted into heat to account for the measured resonance shift induced by the thermo-optical effect. The robustness of the method is guaranteed by a novel procedure for characterizing the thermal time constant in the nanophotonic resonator. The procedure is based on capturing the resonator response to optical intensity modulation across the range of frequencies covering the thermal response timescales. Conceptually, the known thermal time constant simplifies the calculation of the heat generation from light, reducing it from the hard question of thermal impedance modeling to a simpler calculation of the heat capacity of the resonator structure. Therefore, we can obtain the heat dissipation with high confidence, which makes the method more robust and generally applicable. Based on this method, we quantify losses separately for a total of ten whispering gallery modes (WGMs) of three different radial orders with drastically distinct quality factors. The small absorption is found to be nearly identical across devices with order-of-magnitude differences in the total dissipation rate, consistent with possible bulk material absorption. Radiation is found to be the dominant source of loss, and comparing radiation losses across the ten measured modes shows that first- and second-radial-order modes radiate energy mainly via surface scattering, while the third-radial-order one additionally exhibits significant leakage. Using only the response to a single optical stimulus, the simple techniques could be applied to quantitatively separate absorptive and non-absorptive losses in a variety of photonic cavities subject to the ubiquitous self-heating and thermo-optical tuning. Quantitative characterization of both the thermal time constants and the self-heating are important for broadband sensing[23], optical thermometry[24,25] and in other photonic cavity applications using the thermal effect[2–12].

## Measurement of whispering gallery modes

The nanophotonic resonator under investigation is a silicon microdisk which supports WGMs. As shown in Figure 1, the microdisk is a part of a photonic AFM probe that has been demonstrated in various measurements, presenting outstanding sensitivity and bandwidth[23,26]. The microdisk, supported by a silica post underneath at the center, is of 10 μm diameter and 260 nm thickness. A mechanical doubly clamped cantilever probe of 150 nm width surrounds the microdisk with a 200 nm gap. The cantilever is sufficiently narrow and far enough from the disk, so that it causes only a small perturbation to the optical modes. Conducting these experiments on an optomechanical probe instead of a bare microdisk cavity not only demonstrates the use of the technique with a practical nanophotonic sensor, but also enables an additional, independent experimental verification for the thermal time constant. The mechanical cantilever is used to apply frequency modulation to the disk via the optomechanical coupling, providing an alternative way to extract the thermal time constant, to validate the one obtained via the more universally applicable electro-optic intensity modulation approach developed in this work.

A continuous-wavelength tunable laser working from 1510 nm to 1570 nm is used as the light source for measuring the WGMs. The laser light coupled into a single-mode fiber passes through an optical isolator (OI), a variable attenuator (VA), a polarization controller (PC1), and is connected via one of the two paths shown in Fig. 1. For characterizing the thermal time constant, the light is connected to an electro-optic modulator (EOM) and another polarization controller (PC2), eventually coupled to the microdisk via an on-chip waveguide. For swept-wavelength WGM spectra measurement, the light is directly coupled to the microdisk via path 2. The transmitted light is collected by a photodetector (PD) and recorded by an oscilloscope (OSC) and a lock-in amplifier. The device is measured in air at room temperature.

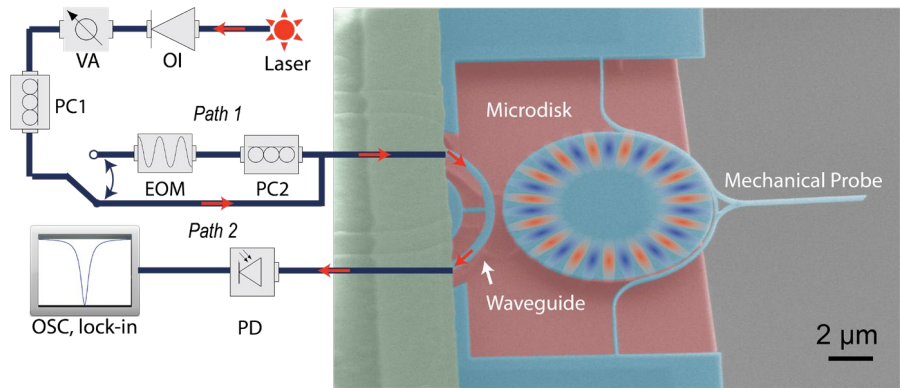

Figure 1 Schematic of the measurement setup. A 10 μm silicon microdisk serves as the nanophotonic resonator under research. The injected laser light passes path 1 or 2 in the experiment of the thermal time constant and dissipation analysis, respectively. The laser light is coupled to the microdisk via an on-chip waveguide to excite the WGMs. The transmitted light is collected by a photodetector. The blue and red microdisk overlay schematically illustrated a WGM.

Figures 2(a) and 2(b) show the WGM resonance dips in the transmission spectrum of the device measured with low and high input powers, respectively. The light passes path 2, and directly couples to the microdisk. While the disk supports both transverse electric (TE) and magnetic (TM) WGMs, the input polarization was adjusted and the integrated waveguide separation was optimized for coupling to $TM_{n,m}$ modes of lowest radial orders $n$ = 1, 2 and 3 [Figure 2(a), inset], identified by their free spectral range (spacing between modes) which are ≈ 14.7 nm, 15.3 nm, and 16.4 nm in the detected frequency range. Note, in the vertical direction of the thin microdisk, we can only measure the lowest order modes (only one node in the vertical direction) as shown in the inset of Fig. 2(a). The modes exhibit distinct quality factors, ranging from ≈ $8.7\times 10^3$ to ≈ $111.1\times 10^3$, as discussed later. A single transmission dip is shown in detail in Fig. 2(c). With increasing light intensity, the optical response exhibits nonlinear behavior, shown in Fig. 2(b), (d). Changing the direction of the laser wavelength scan makes the optical modes show hysteresis, shown in Fig. 2(d).

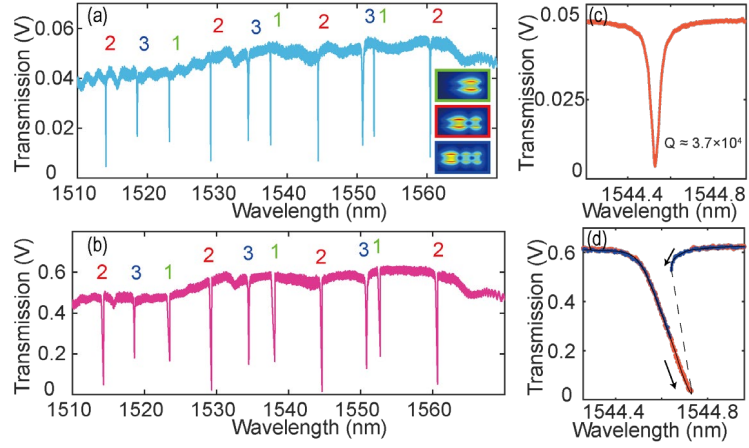

*Figure 2* Wavelength scan of the TM modes. Linear and thermo-optically nonlinear transmission spectra are shown in (a) and (b). The colored numbers label the mode families of distinct radial orders. Their radial-cross-sectional mode-shapes, i.e. the distribution of electric field intensity, are shown in the inset. One optical resonance is shown in detail in (c) and (d), for low and high input power, respectively. The hysteresis in (d) is obtained by changing the direction of the laser wavelength scan, as labeled by the black arrows. The solid (dashed) line labels the stable (unstable) state of the hysteresis. It is fitted based on that the wavelength shift is proportional to the energy absorption in the microdisk, as we will discuss later.

The nonlinearity in the transmission dip for such a small micrometer-scale cavity is due to the thermo-optical effect[1]. The temperature variation $\Delta T$ in the cavity as a function of transmitted light intensity ($\propto$ input optical power $P_{in}$) tunes the resonance wavelength $\lambda_0$ by $\Delta \lambda$ by changing the refractive index $n_0$ and the size of resonator :

$$\frac{\Delta \lambda}{\lambda_0} = \left(\chi + \frac{1}{n_0}\frac{dn}{dT}\right)\Delta T \quad (1)$$

where $\frac{dn}{dT} = 1.72 \times 10^{-4}$ K$^{-1}$ is the temperature sensitivity of the refractive index[27] at the measured wavelength, $\chi = 2.6 \times 10^{-6}$ K$^{-1}$ is the thermal expansion coefficient of silicon, $n_0 \approx 3.48$ for silicon at the working wavelength $\lambda_0 \approx 1.5$ µm at room temperature. As $\frac{1}{n_0}\frac{dn}{dT} \gg \chi$ at the working wavelength and working temperature, Eq. (1) can be simplified to:

$$\frac{\Delta \lambda}{\lambda_0} = \alpha \Delta T \quad (2)$$

where $\alpha = \frac{1}{n_0}\frac{dn}{dT} \approx 4.94 \times 10^{-5}$ K$^{-1}$. The relative wavelength shift is proportional to the temperature change of the cavity which gives us a window to study thermodynamics in the nanophotonic resonators under thermo-optical nonlinearity.

**Total input optical power to the microdisk**

We first need to quantify the total input power to the microdisk. To calculate the input optical power, we characterize the coupling loss between the optical fiber and the on-chip waveguide for both the input and the output. The input and output efficiencies are characterized experimentally by measuring the input and corresponding output powers for light traveling in both directions through the waveguide (i.e., from "terminal 1" to "terminal 2" and vice-versa), as illustrated in Figure 3 (a).

The input and output power coupling ratios from the fiber to the on-chip waveguide at the two terminals are $C_1$ and $C_2$, respectively, which also include the on-chip waveguide losses, if any, to the location of the microdisk. The transmission ratio of the waveguide at the microdisk junction is $C$. At wavelengths away from cavity resonance, the light does not enter the microdisk, resulting in unity transmission $C = 1$, while at the optical resonance $C$ depends on the wavelength.

Accordingly, at wavelengths far from the resonance we have

$$\frac{P_{out}^2}{P_{in}^1} = \frac{P_{out}^1}{P_{in}^2} = C_1 C_2, \qquad (3)$$

where $P_{in}^i$ and $P_{out}^i$ represent the input and output power from "terminal $i$", labeled in Fig. 3(a).

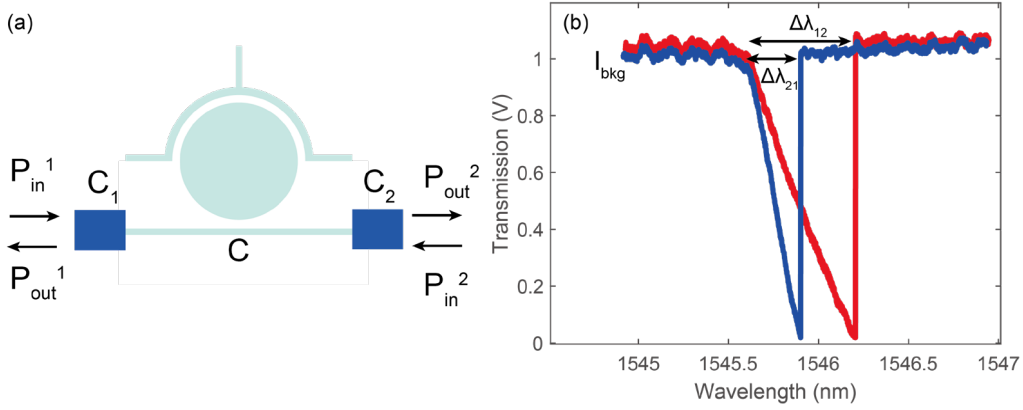

Figure 3 Characterization of the input/output loss of the waveguide. (a) schematic of the microdisk-waveguide system. (b) Measured transmission signal from each end of the waveguide. The different wavelength shift is due to different coupling losses at each end.

At the resonance wavelength with near-optimal waveguide-disk coupling[28], the loss in the microdisk is nearly 100% and $C \approx 0$, i.e. no light passes the junction reaching the output. Based on that, the corresponding resonance shifts $\Delta\lambda$ due to thermo-optical effect are proportional to the power input to the microdisk at the waveguide-microdisk junction, $P_{in}^i C_i$. Therefore, we have:

$$\frac{\Delta\lambda_{12}}{P_{in}^1 C_1} = \frac{\Delta\lambda_{21}}{P_{in}^2 C_2} \qquad (4)$$

where $\Delta\lambda_{i,j}$ are the resonance wavelength shifts of the optical dip [Fig. 3(b)] when the input and output are $i, j$, respectively. From Eq. (3) and (4) we obtain:

$$\begin{cases} C_1 = \dfrac{1}{P_{in}^1}\sqrt{\dfrac{P_{out}^2 P_{in}^2 \Delta\lambda_{12}}{\Delta\lambda_{21}}} \\[2mm] C_2 = \dfrac{1}{P_{in}^2}\sqrt{\dfrac{P_{out}^1 P_{in}^1 \Delta\lambda_{21}}{\Delta\lambda_{12}}} \end{cases} \qquad (5)$$

The measured input/output powers, $P_{in}^1 \approx 3.583$ mW, $P_{out}^2 \approx 0.014$ mW, $P_{in}^2 \approx 3.586$ mW, $P_{out}^1 \approx 0.013$ mW give $C_1 \approx 0.0897$ and $C_2 \approx 0.0425$. In subsequent experiments, terminal 1 is used as the light input, therefore, the transmission intensity at the waveguide-microdisk junction is $\frac{I_{bkg}}{C_2}$, where $I_{bkg}$ is the measured transmission intensity at wavelength away from the resonance. The corresponding input power at the waveguide-microdisk junction is written as:

$$P_{in} = \rho \frac{I_{bkg}}{C_2} \qquad (6)$$

where ρ is the gain of the photodetector which is measured to be $\approx 1.44 \times 10^{-5}$ W/V.

**Thermal time constant of nanophotonic resonator**

The origin of the thermo-optical effect is the temperature increase due to the optical energy absorption by the microdisk. As shown in Eq. (2), we can characterize the temperature change of the microdisk by measuring the resonance wavelength tuning of the WGMs. In order to further convert the temperature change into the generated heat from the absorption loss, we need an accurate thermal impedance modeling of the system, including the microdisk, silica post, substrate and surroundings. However, modeling the thermal impedance of a nanophotonic resonator is not easy, especially for those of complex geometry and anchors. Errors in thermal impedance modeling will introduce uncertainty in the later dissipation analysis. On the other hand, a known thermal time constant can directly link the harder-to-model thermal impedance to the resonator's heat capacity, which is well defined by the disk size and the known specific heat capacity of cavity materials, such as Si. Therefore, a direct experimental measurement of the thermal time constant is necessary to obtain the thermal impedance with high confidence. The thermal impedance can then be used to convert the thermo-optically observed temperature change to the absolute absorbed optical power, allowing us to quantify the optical absorption.

Whenever an FEM thermal simulation is used to obtain the thermal impedance, it may be subject to various material and interfacial thermal conductivity uncertainties and potential errors in the computation of heat dissipation. However, one can be more confident in modeling various material heat capacities correctly. Therefore, one needs to run a dynamic time-domain simulation of the system's impulse or step response and compare the simulated thermal time(s) to those obtained experimentally, to verify the correct modeling of the thermal impedance.

To measure the thermal time constant, we switch the connection to the path 1. The light intensity is modulated by an EOM. The light of high intensity couples to the microdisk and excites the thermo-optically nonlinear WGMs. The output light coupled to the thermal dynamics in the microdisk is demodulated by a lock-in amplifier at the modulation frequency.

Under perturbative modulation of intensity, the amplitude of modulated transmission light is proportional to both the modulation magnitude and temperature variation due to the thermo-optical effect. By considering the thermodynamics in the nanophotonic resonator, we obtain the transmission light coupled to the disk (*TM* components) as [See supplementary information]:

$$TM = c_1 + \frac{c_2}{1 + i\omega\tau} \qquad (7)$$

where $\tau$ is the thermal time constant, $\omega$ is the modulation frequency, $c_1$ and $c_2$ are two positive numbers.

Figure 4 shows the measured *TM* after calibration of the propagation delay in the coaxial cable and removing the *TE* component arising from the birefringent effect of the EOM [See supplementary information]. When $\omega \gg 1/\tau$, the thermodynamics cannot follow the modulation speed and the thermal effect is negligible, $TM = c_1$. When $\omega \ll 1/\tau$, the temperature of the disk adiabatically follows the intensity variation, adding another term $c_2$ accounting for the temperature-induced intensity changes, $TM = c_1 + c_2$. We obtain $c_1 \approx 5.5$ mV from the real part of *TM* at $\omega \gg 1/\tau$ and $c_1 + c_2 \approx 22.8$ mV at $\omega \ll 1/\tau$. Only one adjustable parameter $\tau = (6.8 \pm 0.3)$ μs is used to fit the real part of *TM* based on Eq. (7). The uncertainty is the statistical uncertainty of the fit parameter. All uncertainties reported are one standard deviation unless noted otherwise. The corresponding in-phase, out-of-phase, and amplitude of *TM* with the fitted parameters are shown as the black lines in Figure 4, showing excellent agreement.

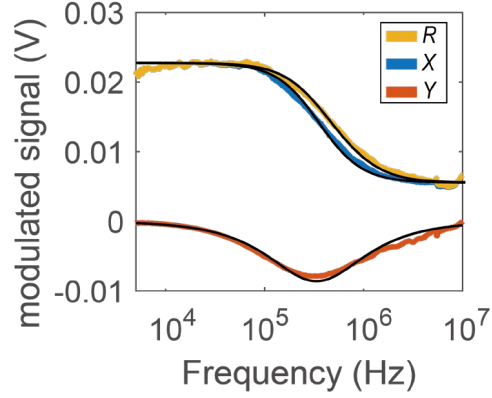

Figure 4 Measured amplitude of modulated *TM* signal. Yellow, blue, and orange lines are amplitude *R*, in-phase component *X*, and out-of-phase component *Y*. The black lines are the fit of Eq. (7)'s amplitude, real and imaginary parts.

The thermal time constant can also be obtained by using frequency (phase) modulation, rather than amplitude modulation. However, the depth of frequency modulation using an EOM phase modulator decreases linearly with the modulation rate, making it difficult to obtain pure frequency modulation without being overwhelmed by the residual intensity modulation at low frequencies. Therefore, the intensity modulation is easier to use for the characterization of the thermal time constant of nanophotonic resonators. However, for our specific broadband optomechanical sensor, an alternative way for obtaining effective frequency modulation is via optomechanical modulation. Recently we have used the cantilever to modulate the frequency of WGMs[18] and obtained $\tau \approx 7.0$ μs, consistent with the one obtained above, further verifies the validity of the method.

**Dissipation analysis**

In the following, we study the thermo-optical nonlinearity of WGMs with different radial numbers and quantitatively separate their absorption loss from other loss mechanisms. Figures 5(a)-(c) present the normalized transmission signal for modes of radial orders of 1, 2, and 3 (family 1, 2, and 3) where the normalized linear and nonlinear responses are colored in blue and purple, respectively. The input power $P_{in}$ at the microdisk-waveguide junction is measured to be $(1.6 \pm 0.1)$ μW and $(163.7 \pm 2.7)$ μW for the linear and nonlinear cases, based on Eq. (6). It is clear that the total dissipation rates for different families are highly distinct, evident from their quality factors. The narrow family-1 resonances show as partially resolved doublets arising from the weak coupling and hybridization between the clockwise (CW) and counterclockwise (CCW) propagating WGMs. Here, we define $\gamma_c$ as the rate of dissipation induced by the coupling loss from the microdisk into the waveguide, $\gamma_a$ is the absorption loss converting optical power into heat and $\gamma_r$ is the radiative dissipation rate of the disk. We write the total dissipation rate $\gamma_t$ of the microdisk cavity as the sum of these terms:

$$\gamma_t = \gamma_c + \gamma_a + \gamma_r \quad (8)$$

Using the conventional waveguide-cavity-coupling model [28,29], we write the normalized

linear transmission intensity as [19]:

$$I = \frac{\left|-\sqrt{P_{in}} + \sqrt{\gamma_c/2}(a_1 + a_2)\right|^2}{P_{in}} \quad (9)$$

where $a_{1,2} = \frac{-\sqrt{\gamma_c/2}\sqrt{P_{in}}}{-(\gamma_t/2)+i\left(-\frac{2\pi c}{\lambda_0^2}\lambda_\Delta \pm \gamma_\beta/2\right)}$ are the amplitudes of the doublet standing wave modes. $\lambda_\Delta$ is the detuning of laser wavelength from the optical mode, $c$ is the speed of light, and $\gamma_\beta$ is the coupling rate between the CW and CCW traveling wave modes, giving rise to the doublet split (its value is defined by the size of the split in the spectrum). For families 2 and 3 where the doublet is unresolved, $\gamma_\beta$ is set to 0 Hz. Noting that our system is in the undercoupled regime [28,30], fitting the linear transmission signal yields $\gamma_t$ and $\gamma_c$ for each mode, as shown in the Table 1, and the yellow and blue bars in Fig. 5(d). Note, we use the least square fit to extract $\gamma_t$ and $\gamma_c$ in Eq. (9), other parameters such as $P_{in}$, $\lambda_\Delta$, and $\lambda_0$ are obtained individually prior to fitting. It is evident that the total dissipation rate increases with radial numbers, due to the increased radiative loss rates, as we show below.

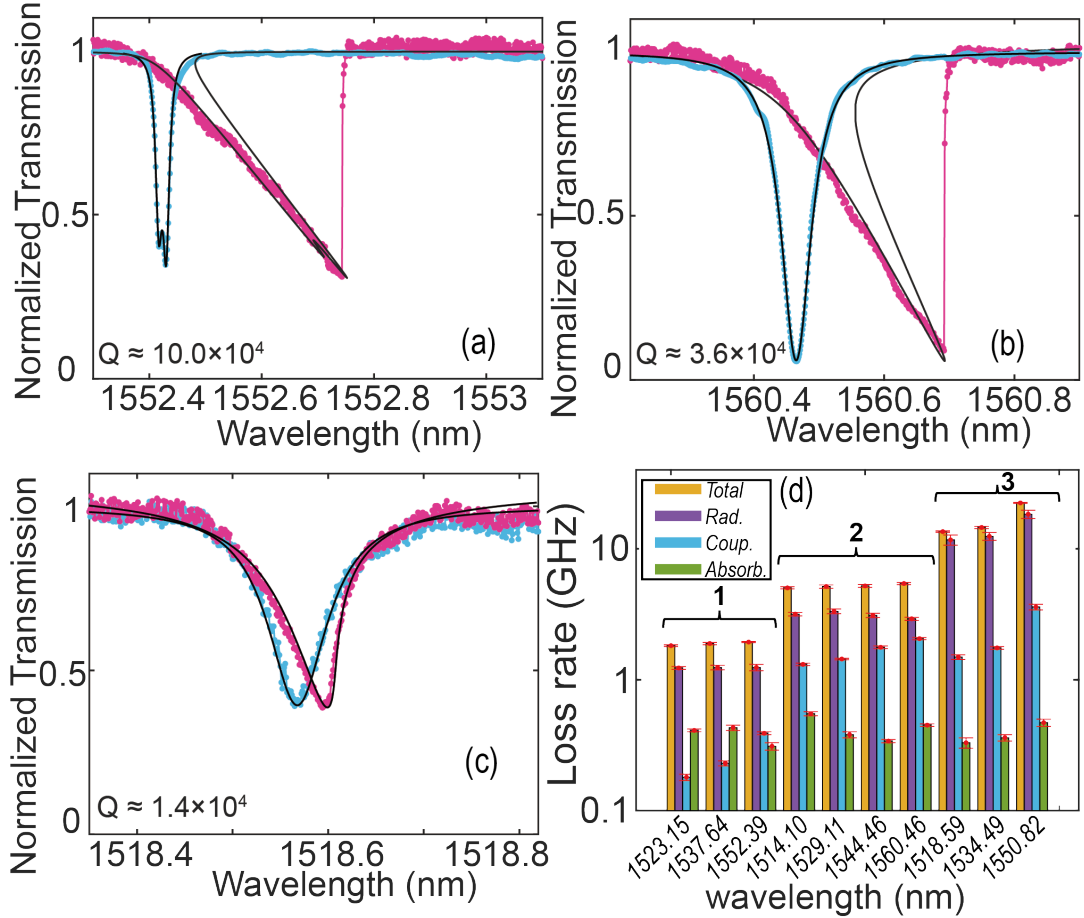

*Figure 5.* Thermo-optical nonlinearity for WGMs of distinct radial numbers. Normalized linear (blue) and nonlinear (purple) transmission signals of the modes with the radial orders of 1, 2, 3 are shown in (a), (b) and (c). The black lines are the corresponding fits from Eq. (9) and (10), respectively. The input power $P_{in}$ before the microdisk is measured to be around 1.6 μW and 163.7 μW for the linear and nonlinear cases. (d) Distinguished loss rates of different modes. Yellow, purple, blue and green bars (arranged left to right for each mode) show the total, radiative, coupling, and absorptive loss rates, respectively. The numbers on the top label the corresponding radial family. The modes are grouped by their radial order as in Table 1. The error bars are one standard deviation statistical uncertainties propagated from fit parameters

| Radial order | Resonance wavelength (nm) | Loss rate (GHz) | | | |
|---|---|---|---|---|---|
| | | Total | Coupling | Absorption | Radiation |
| 1 | 1523.15(2) | 1.82(3) | 0.18(1) | 0.41(1) | 1.23(3) |
| | 1537.64(1) | 1.89(4) | 0.23(1) | 0.43(2) | 1.24(5) |
| | 1552.39(2) | 1.94(2) | 0.39(1) | 0.31(2) | 1.24(7) |
| 2 | 1514.10(2) | 5.03(9) | 1.31(3) | 0.55(2) | 3.17(10) |
| | 1529.11(2) | 5.15(15) | 1.44(2) | 0.38(2) | 3.33(14) |
| | 1544.46(1) | 5.21(13) | 1.77(4) | 0.34(1) | 3.10(12) |
| | 1560.46(1) | 5.43(11) | 2.06(4) | 0.45(1) | 2.92(8) |
| 3 | 1518.59(1) | 13.52(18) | 1.49(6) | 0.33(3) | 11.70(104) |
| | 1534.49(2) | 14.55(31) | 1.75(4) | 0.36(2) | 12.45(85) |
| | 1550.82(1) | 22.42(20) | 3.59(18) | 0.47(3) | 18.37(134) |

*Table 1* Separated loss rates for different modes of radial orders 1, 2 and 3. Uncertainties in the last significant digit are shown in parenthesis, which are one standard deviation statistical uncertainties propagated from fit parameters.

To distinguish the absorption loss from radiation loss in the microdisk, we consider the high-power spectrum of the mode. Different from the method shown in Ref. [19], which determines the absorption loss by analyzing strong nonlinear absorption, here, we determine the absorption loss by calculating the amount of power that is being converted into heat to account for the measured resonance shift. The proposed method is generally applicable to photonic resonators subject to thermo-optical nonlinearity, does not rely on material optical nonlinearities, and is capable of analyzing the dissipation mechanisms of optical modes of different radial orders with very different loss rates.

Based on Eq. (2) $\frac{\Delta\lambda}{\lambda_0} \approx \alpha\Delta T = \alpha\kappa Q_a$, we rewrite the relative resonance frequency shift as:

$$\frac{\Delta\lambda}{\lambda_0} \approx \alpha\kappa\sigma(P_{in} - P_{out}) \quad (10)$$

where $P_{out}$ is the output power at the waveguide-disk junction, $\kappa = 1.50 \times 10^5$ K/W is the temperature change of the disk per unit of absorbed power obtained from the finite-element-method [see Supplementary Materials], $Q_a = \sigma(P_{in} - P_{out})$ is the power absorbed by the disk and $\sigma = \frac{\gamma_a}{\gamma_a + \gamma_r}$ expresses the ratio of absorbed power over the total optical power loss from the disk. $P_{in}$ is obtained from Eq. (6), and similarly, $P_{out} = \rho \frac{I_{out}}{C_2}$ where $I_{out}$ is the measured output transmission intensity.

Note, we only use one parameter, $\kappa$, from the numerical simulation, which is verified by numerically obtaining a thermal time constant $\tau \approx 6.8$ µs, in agreement with the experimental value. Since the time constant $\tau$ couples the thermal impedance $\kappa$ with the microdisk heat capacity, which only depends on the geometric size and the material's specific heat capacity of the microdisk, the correct time constant verifies the value for $\kappa$ obtained by the numerical model.

Expressing power through the normalized measured transmission, $(P_{in} - P_{out}) = \frac{\rho}{C_2}(I_{bkg} -$

$I_{out}) = \frac{\rho}{C_2}\eta(1-I)$ with a normalization coefficient $\eta$ and using Eq. (10) and (9), we fit the normalized nonlinear transmission signal shown as the black lines in Figure 5. The fitting procedure is summarized in three steps. First, we fit the linear resonance under low input power using Eq. (9), shown as the blue colored peaks and their fit in Fig. 5. Next, we shift the wavelength of the linear resonance as a function of the absorption energy in the cavity $\sigma(P_{in} - P_{out})$ as shown in Eq. (10). Last, by sweeping the value of $\sigma$ we fit the thermo-optically nonlinear resonance based on the least square fit. There is only one fitting parameter $\sigma = \frac{\gamma_a}{\gamma_a + \gamma_r}$. From $\sigma$ and the known $\gamma_t$ and $\gamma_c$, we obtain the $\gamma_a$ and $\gamma_r$. They are shown in Table 1 and the green and purple bars in Fig. 5(d).

The absorption losses are found to be similar across all the investigated highly-confined modes, favoring the bulk absorption of silicon as the root cause. The uniform absorptive loss rate makes the optical modes of a low quality factor show less thermo-optical nonlinearity for the same input power since the absorption loss is a relatively smaller portion of the total loss. It means that low-quality-factor modes possess a larger power dynamic range before entering nonlinearity where thermal-instability and nonlinear dynamics show up.

As expected, the waveguide coupling rates are found to increase systematically with higher radial and lower azimuthal numbers, as the WGM evanescent tail increases and more strongly overlaps with the waveguide mode. While they also depend on the effective index of the waveguide, the leading trend is the coupling increase with the lowering of the mode effective index inside the microdisk and the corresponding lengthening of the evanescent tail.

The radiation losses are the dominant losses for all modes and are larger for the higher radial order families. As shown in Figure 6, for first and second radial order families, the radiation loss does not significantly depend on the azimuthal number, showing the loss is mainly from scattering loss, presumably due to surface roughness and point defects. The surface scattering notably increases with the radial number in agreement with theoretical predictions[31]. Additionally, for the modes of the third radial order, the radiation loss increases with decreasing azimuthal number and is linearly correlated with the waveguide coupling loss, suggesting that the radiation leakage plays a significant role in addition to scattering due to imperfections. As the radiation losses exponentially increases with decreasing the disk size[32], we expect the radiation loss remains as the dominant loss mechanism for smaller disks.

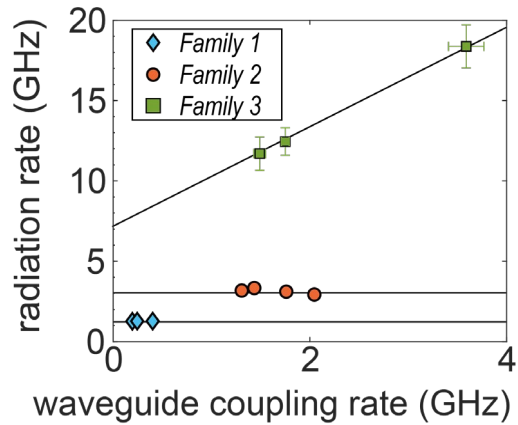

Figure 6. Radiative loss rate plotted vs. waveguide coupling loss rate. The blue diamonds, orange circles and green squares correspond to radial orders 1, 2, and 3, respectively. The black lines are drawn to guide the eye. The error bars are smaller than the size of marks for families 1 and 2.

## Conclusions

We systematically analyze the dissipation of WGMs with various radial and azimuthal orders. By quantitatively characterizing the thermal time constant of the system, we obtain a better understanding of thermo-optical dynamics in the system, which enables us to perform a systematic study and disambiguation of the absorption and radiation optical loss mechanisms in the optical cavity modes, including their dependence on mode frequency and radial structure. For the microdisk resonators studied here, the loss is dominated by scattering due to surface roughness, with additional radiative leakage loss for the high-radial-order mode family. The uniform absorption losses, independent of mode radial number, favor bulk over surface-localized absorption mechanism. The described method can be used to disambiguate loss mechanisms in other photonic cavities subject to thermo-optical nonlinearity. The novel method to characterize the thermal time constant and self-heating can benefit optical thermometry, photonic sensing and other photonic cavity applications.

**Acknowledgments**
We thank Dr. Jeffrey Schwartz, Dr. Biswarup Guha, Dr. J. Alexander Liddle and Dr. Marcelo Davanco for reviewing this paper and giving meaningful suggestions. M.W. and D.J.P. are supported by the Cooperative Research Agreement between the University of Maryland and the National Institute of Standards and Technology Center for Nanoscale Science and Technology, Award 70NANB10H193, through the University of Maryland.

**Author contributions**
V.A. conceived the project. M.W. and D.J.P designed the research. D.J.P. and M.W. fabricated the devices. M.W. and D.J.P. performed the experiments. M.W. analyzed the data. All authors discussed the results and contributed the manuscript.

**Data availability**
The data that support the findings of this study are available from the corresponding author on request.

**Conflict of interest**
The authors declare no competing interests.

# Supplementary Materials

**Extraction of thermal time constant via intensity modulation**

As thermal time constants in nanophotonic resonators may be several microseconds or longer, frequency modulation via phase electro-optical modulators (EOM) requires deep phase modulation that is challenging to implement. Therefore we implement a method that extracts the thermal time constant based on much more easily achievable intensity modulation,. The method relies on the interplay between the optical response and the thermodynamics in nanophotonic resonators via thermo-optical effect. As shown in Figure S1 (a), if the input optical intensity is reduced by $\Delta I_0$, the transmission intensity at the resonance (red dot) will first reduce to the scaled solid blue line quickly, and then slowly decay to the dashed blue line due to the thermo-optical tuning of resonance. Note, Fig. S1 (a) exaggerates the intensity modulation for a better presentation, the actual modulation is just a perturbation ($\Delta I_0 \ll I_0$). The tuning-induced variation of the transmission can be simply understood, as that the decrease of intensity lowers the temperature in the resonator, shifting the resonance wavelength by $\Delta\lambda$ to a lower wavelength. The resonance shift leads to extra intensity change proportional to $\Delta\lambda$. Furthermore, $\Delta\lambda$ is proportional to $\Delta T$ based on Eq. (2). Considering the two parts, we have:

$$\Delta I = k\Delta I_0 + p\Delta T \qquad (S1)$$

where $\Delta I$ is the intensity variation at the resonance due to intensity modulation $\Delta I_0$, $k$ and $p$ are two proportionality constants, $\Delta T$ is the temperature variation in the nanophotonic resonator and it can be expressed by the thermodynamic equation:

$$\frac{d(\Delta T)}{dt} = -\frac{1}{\tau}(\Delta T - \Delta I_0) \qquad (S2)$$

where $\tau$ is the thermal time constant and we express the temperature change $\Delta T$ in the units of intensity change $\Delta I_0$ for simplicity of notation ($\Delta T = \Delta I_0$ defines the steady state relationship of temperature on intensity change). Finally, we write the intensity modulation as harmonic drive, and the resulting intensity and temperature changes are also in the harmonic form:

$$\begin{aligned}\Delta I_0 &= \widehat{I_0} e^{i\omega t}\\ \Delta I &= \hat{I} e^{i\omega t}\\ \Delta T &= \hat{T} e^{i\omega t}\end{aligned} \qquad (S3)$$

By applying Eq. (S3) to (S1) and (S2), we obtain the amplitude $\hat{I}$ as:

$$\hat{I} = \widehat{I_0}\left(k + \frac{p}{1 + i\omega\tau}\right) \qquad (S4)$$

At $\omega \gg 1/\tau$, the thermodynamics cannot follow the quick modulation, so the thermal tuning effect

(second term in Eq. S1) is negligible, $\hat{I} = k\hat{I}_0 = c_1$. At $\omega \ll 1/\tau$, the resonator is always at equilibrium, therefore, the amplitude of intensity is affected by the thermal effect: $\hat{I} = k\hat{I}_0 + p\hat{I}_0 = c_1 + c_2$. Since $c_1$ and $c_2$ can be measured individually, Eq. (S4) is simplifier to:

$$\hat{I} = c_1 + \frac{c_2}{1 + i\omega\tau} \tag{S5}$$

In practice, rather than using a dedicated amplitude-modulating EOM, we have used a phase EOM to achieve amplitude modulation. This was possible due to the combination of the EOM's polarization-dependent phase modulation strength and the polarization-dependent fiber to photonic waveguide coupling. When the EOM input polarization is deliberately misaligned from the EOM principal axes and the EOM polarization principal axes are in turn misaligned from the on-chip waveguide input coupler axes, the TE and TM modes in the waveguide both acquire amplitude modulation. Only *TM* mode is coupled to the optical cavity modes under consideration, therefore the *TE* component is just a constant modulated intensity background independent from $\omega$. Moreover, due to energy conservation, the intensity of *TE* is out of phase from the *TM* signal, providing a negative background $-b_1$ to the in-phase components of the phase sensitive readout. Therefore, the overall readout is:

$$\hat{I} = B_1 + \frac{c_2}{1 + i\omega\tau} \tag{S6}$$

where $B_1 = c_1 - b_1$.

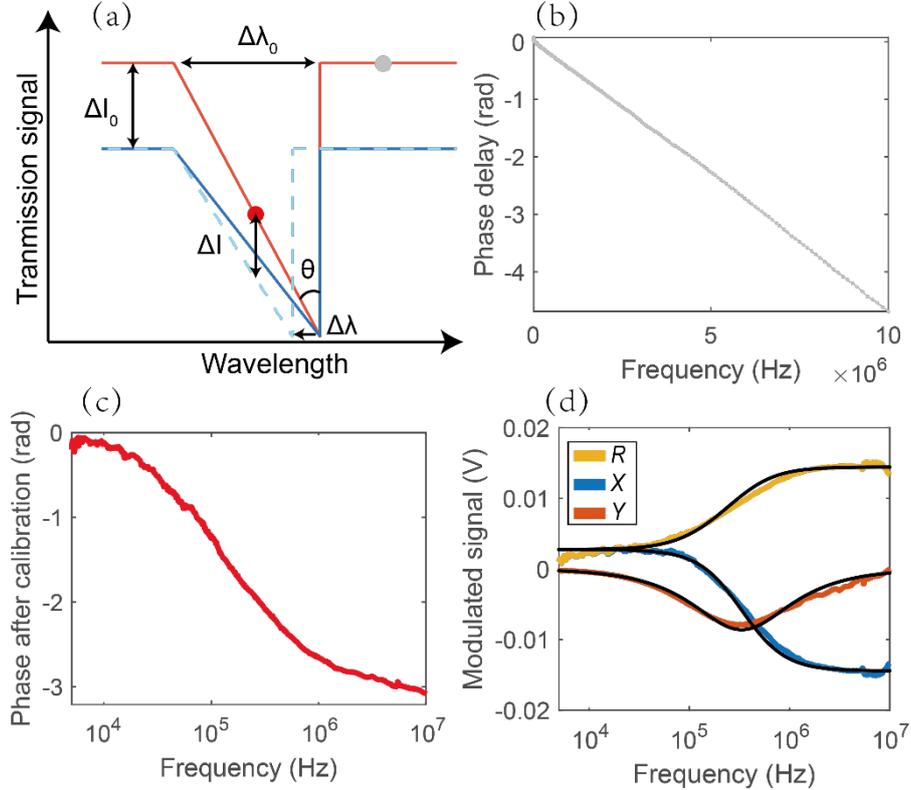

Figure S1. Characterization of the thermal time constant via intensity modulation. (a). Schematic of transmission signal under the intensity modulation. The change of transmission intensity at the working wavelength (red dot) contains two parts which are from the red line to the solid blue line and further to the dashed blue line. (b) Phase as a function of modulation frequency measured at the background (gray dot). (c) Calibrated phase as a function of modulation frequency measured at the working wavelength. (d) Measured modulated transmission intensity. Yellow, blue, and orange lines are amplitude $R$, in-phase component $X$, and out-of-phase component $Y$. The black lines are the fit of Eq. (S5)'s amplitude, real and imaginary parts.

In the experiment, we first measure the phase delay induced by the electric transmission line at the wavelength of the background (gray dot), shown as Fig.S1 (b), and use it to calibrate the phase of the signal measured at the working wavelength (red dot), shown as Fig.S1 (c). The amplitude, in-phase and out-of-phase parts of the calibrated signal are presented in Fig.S1 (d) as the yellow, blue, and orange dots, respectively. At the frequency of around 10 MHz, we get $B_1 \approx -14.5$ mV from the in-phase component, while at around 10 kHz we get $B_1 + c_2 \approx 2.8$ mV. With the known $B_1$ and $c_2$, we fit the in-phase component by Eq. (S6), and obtain $\tau = 6.8 \pm 0.3$ µS. To better present the *TM* part of the signal, we exclude the *TE* components by assuming *TM* and *TE* are nearly equal at low frequencies, so $b_1 \approx -20$ mV. This assumption does not affect the fitting procedure at all. In Figure 4, we add 20 mV to the in-phase component to exclude the *TE* part for a better presentation of the *TM* signal.